\documentclass[twocolumn,showpacs,preprintnumbers,amsmath,amssymb]{revtex4}

\usepackage{graphicx}
\usepackage{dcolumn}
\usepackage{bm}
\usepackage{epsfig}
\epsfclipon

\newcommand{\nco}{\newcommand}
\nco{\beq}{\begin{equation}} \nco{\eeq}{\end{equation}}
\nco{\beqa}{\begin{eqnarray}} \nco{\eeqa}{\end{eqnarray}}
\nco{\lra}{\leftrightarrow}

\nco{\sss}{\scriptscriptstyle} \nco{\dphi}{\varphi}
\nco{\lsim}{\mbox{\raisebox{-.6ex}{~$\stackrel{<}{\sim}$~}}}
\nco{\gsim}{\mbox{\raisebox{-.6ex}{~$\stackrel{>}{\sim}$~}}}

\def\pref#1{(\ref{#1})}

\begin{document}

\preprint{McGill 03-25}

\title{The phantom menaced: constraints on low-energy effective ghosts}

\author{James M.\ Cline, Sangyong Jeon, Guy D.\ Moore}

\affiliation{%
\centerline{Physics Department, McGill University,
3600 University Street, Montr\'eal, Qu\'ebec, Canada H3A 2T8}
e-mail: ghostbusters@physics.mcgill.ca }

\date{November, 2003}

\begin{abstract}
It has been suggested that a scalar field with negative kinetic energy,
or ``ghost,'' could be the source of the observed late-time cosmological
acceleration. Naively, such theories should be ruled out by the
catastrophic quantum instability of the vacuum.  We derive
phenomenological bounds on the Lorentz-violating
ultraviolet cutoff $\Lambda$ which must
apply to low-energy effective theories of ghosts, in order to keep the
instability at unobservable levels.  Assuming only that ghosts interact
at least gravitationally, we show that $\Lambda \lsim 3$ MeV for consistency
with the cosmic gamma ray background.  We also show that theories of
ghosts with a Lorentz-conserving cutoff are completely excluded.
\end{abstract}

\maketitle

The present accelerated expansion of the universe seems to be an
experimental fact, now that data from distant type Ia supernovae
\cite{SCP} have been corroborated by those from the cosmic microwave
background \cite{WMAP}.  Although the simplest explanation is a
cosmological constant $\Lambda$ of order $(10^{-3}$ eV)$^4$, this tiny
energy scale is so far below the expected ``natural'' size for a
cosmological constant, that
alternative explanations have been vigorously pursued.  A common
approach has been to assume that the true value of $\Lambda$ is zero,
due to an unknown mechanism, and to propose new physics which would
explain why the present-day vacuum energy differs from zero by the small
observed amount.  

The most popular idea has been quintessence, in which the universe is
gradually approaching the zero of the vacuum energy by the slow rolling
of an extremely weakly coupled scalar field.  More recently, some less
conventional alternatives have been considered, including ``phantom
matter,'' which is essentially quintessence with a wrong-sign kinetic
term \cite{phantom}. These models are motivated by the supernova data,
which suggest that the dark energy equation of state violates the weak
energy condition by having $p<-\rho$ \cite{WEC}.

A serious problem with phantom matter, which is overlooked in the literature
that attempts to apply it to cosmology, is that
such theories are not quantum mechanically viable, either  because they
violate conservation of probability, or they have unboundedly 
negative energy density and lead
to the absence of a stable vacuum state.  Whether a ghost carries negative
norm and positive energy, or vice versa, is a choice which is made during
the quantization procedure.  This choice exists because the $i\epsilon$
prescription for defining the propagator near its poles is not unique,
and not specified by the Lagrangian itself.  The momentum space propagator
for a ghost can have either of the two forms
\beq
\label{prop}
	{-i\over p^2-m^2 +i\epsilon} \quad\hbox{\ or \ } 
	{-i\over p^2-m^2 -i\epsilon} 
\eeq

In the first form in \pref{prop}, the imaginary part of the propagator has
the opposite sign relative to that of a positive norm particle.  This will
cause the optical theorem to be violated, leading to a nonunitary
theory.  That is, this choice gives a theory with no probabilistic
interpretation.  It is therefore unphysical and should be dismissed.

%

On the other hand, if the second form in \pref{prop} is chosen, unitarity is
maintained. The price to be paid is that the poles in the propagator are
shifted in such a way that particles with negative energy are the ones which
propagate forward in time, so ghosts possess negative energy.  This
means, for instance, that 
a two-body scattering process involving
nonghosts and ghosts can result in an increase in the magnitude of
the energies of the particles.  To illustrate this, suppose that the ghost is
massive so that we  can consider it to be initially at rest. If the initial
energy of a photon is $E_i$ and it gravitationally scatters from the ghost 
at angle $\theta$, then its final energy is 
\beq
	E_f = E_i \left({m\over m -E_i(1-\cos\theta)} \right) 
	> E_i
\eeq
in contrast to the nonghost case where photons can only lose energy in such
scatterings.
In fact, there exist initial energies $E_i = m/(1-\cos\theta)$ such that
the final energy is divergent.  The final energy of the ghost is
correspondingly large and negative.  

To avoid this kind of problem, one should consider theories where the
interactions between ghosts and normal matter are as weak as possible.
However, we must allow
the ghosts to interact gravitationally, since it is their gravitational
interactions which are needed for them to have any cosmological
consequences, and this is already enough.  Gravitational interactions
allow the process of figure 1, in which a ghost pair and photon pair are
spontaneously created from the vacuum.
The phase space integral is divergent, indicating a catastrophic
instability.  The divergent nature of the instability can only be
avoided if we impose a {\em Lorentz noninvariant} momentum space cutoff
on the final state phase space (more on this later).  
Setting such a cutoff at the scale
$\Lambda$, the creation rate is, on dimensional grounds,
\beq
\label{rate}
	\Gamma_{0\to 2\gamma 2\phi} \sim {\Lambda^8\over M_p^4} \, .
\eeq
We have neglected Bose enhancement from final state occupancy, an
assumption we will verify {\em a posteriori}.
Notice that this pathology exists independently of the question of classical
stability of the ghost-gravity system, which has been considered in
\cite{kam}.  The existence of this process
is model independent---it requires only the existence of a wrong sign,
canonical kinetic term and gravitational interaction for the ghosts.

\smallskip \centerline{\epsfxsize=0.2\textwidth\epsfbox{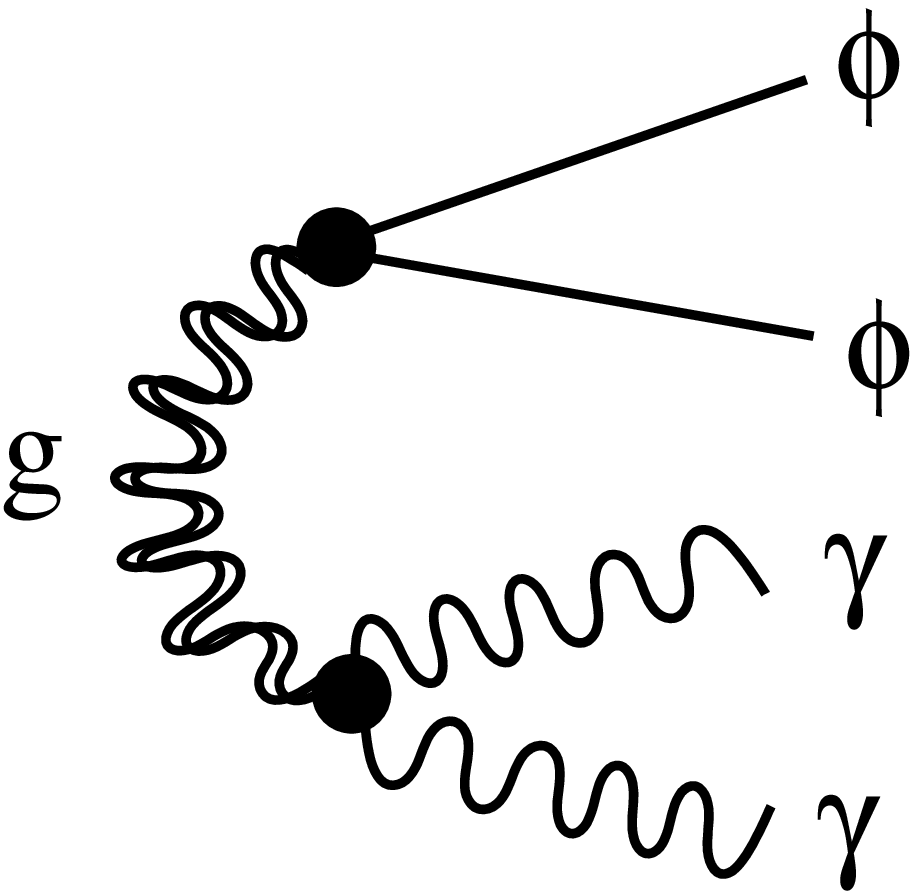}}
\begin{verse}
{\small Figure 1. Graviton-mediated decay of vacuum into two ghosts and
two photons.}
\end{verse}

An implicit excuse for even considering phantom matter at the classical
level is perhaps the idea that, at long distances, the scalar field
theory is merely an 
effective one, whose ultraviolet completion is well defined and respects
unitarity.  In this way, it might be possible to have a physical value for the
cutoff in  \pref{rate}  which was small enough so that the rate of decay of the
vacuum is slow on cosmological time scales.  In this letter, we
estimate just how low a cutoff $\Lambda$ is required for consistency with
observational constraints.  This question was previously considered in
\cite{Carroll}; but we reach somewhat different conclusions, as we
discuss below.

To find the density of
photons which are spontaneously produced, we evolve the phase space
density of ghosts and photons in an expanding universe,
\beq
\label{Boltz}
	{d\over dt}\left(a^3\, n\right) = a^3\, \Gamma \, ,
\eeq
where $a(t)$ is the scale factor and $\Gamma=\Gamma_{0\to 2\gamma 2\phi}$.  The
solution is
\beq
\label{den}
	n(t) = \Gamma  \left\{\begin{array}{ll} {t\over 3p+1},&
	a(t) \sim t^p\\ H^{-1}(1-e^{-3Ht}),& a(t)\sim e^{Ht} \end{array}\right.
	\, .
\eeq
That is, the current number density is approximately given by the
production rate per spacetime volume, $\Gamma$, times the age of
the universe.  Most of the photon pairs have been produced since
redshift $z=1$, both because there has been more time since $z=1$ than
before then, and because the density produced earlier was diluted by the
expansion of the universe.  This also means that their energy spectrum
is not very different from the energy spectrum produced today; the
spectrum peaks at $E \sim \Lambda$.  Therefore we find,
\beq
\label{spect}
	{dn\over dE} \sim \Lambda^7 M_p^{-4} t_0 \quad
	\mbox{for\ }E \lsim \Lambda \, .
\eeq

This spectrum of photons with energy near $\Lambda$ is constrained by
observations of the diffuse gamma ray background.  EGRET \cite{egret}
has measured the  differential photon flux to be 
\beq
\label{dFdE2}
	{dF\over dE} = 7.3\times 10^{-9}\left({E\over E_0}\right)^{-2.1}
	\left(\hbox{\ cm}^2\hbox{\ s\ sr\ MeV}\right)^{-1} \, ,
\eeq
where $E_0 = 451$ MeV.  Demanding that \pref{spect} not exceed \pref{dFdE2} 
gives the upper limit
\beq
\label{constraint}
	\Lambda \lsim 3 \hbox{\ MeV} \, .
\eeq
Since the observed gamma ray spectrum involves a mean particle occupancy
which is orders of magnitude less than 1, neglect of Bose stimulation
was entirely justified.

We emphasize that this bound depends only upon the ghost having at least
a minimal coupling to gravity.  The possible presence of other couplings
can only strengthen the result. 
Nor does it depend on whether the ghost has a potential, so long as its mass is less than
$\Lambda$.  In models of phantom cosmology, the mass is taken to be of order the present
Hubble scale, $10^{-33}$ eV, so this is not restrictive.

The process $0 \to 2\gamma 2\phi$ is not the only allowed one; we
can consider also the production of neutrinos and of $e^+ e^-$ pairs.
The neutrinos are hard to observe, so no good constraint arises there.
The $e^+ e^-$ constraint may be more fruitful, but the existence of
galactic and Earth magnetic fields makes it somewhat more difficult to
relate an incident $e^+$ flux to the extragalactic density.  However, a
sufficiently dense intergalactic $e^+ e^-$ plasma would lead to
excessive rescattering of the cosmic microwave sky.
This leads to the constraint $\Lambda \lsim 40$ MeV, still weaker than
\pref{constraint}.

Let us compare our bound \pref{constraint}
to those which were obtained in ref.\
\cite{Carroll}.  There it was argued that one can constrain $\Lambda <
10^{-3}$ eV by considering the process $\phi\to g\,\phi\,\phi$, where
$g$ is a graviton.  This bound is incorrect, however.  First of all, it
arose by considering the ``decay'' rate of a ghost at rest, and
insisting that this be longer than the Hubble time to prevent the
exponential runaway generation of ghosts.  But the produced ghosts
typically carry energies $\sim \Lambda$, so their decay rates are
strongly time dilated.  Demanding only that the time dilated
decay rate be less than $1/t_0$ gives $\Lambda < 50$ MeV.  
Second, the constraint arose by
considering an interaction ${\cal L}_{\rm eff} = \phi g^{\mu \nu}
\partial_\mu \phi \partial_\nu \phi$.  Not only is this interaction
model dependent; it is actually an 
artifact of a noncanonical normalization of the $\phi$ field kinetic
term, and can be removed by a field redefinition.  Therefore, the
proposed decay mechanism does not actually work.  Diagrammatically,
this is because the total amplitude is zero, as shown in figure
2.  The naive contribution from the contact term is canceled by
the three other diagrams, built from the $\phi\eta^{\mu \nu}
\partial_\mu \phi \partial_\nu \phi$ vertex contained in the interaction,
and the $g^{\mu\nu}\partial_\mu \phi \partial_\nu \phi$ vertex from 
the standard kinetic term. Thus the
rate for $\phi\to g\phi\phi$ is zero.
Ref.\ \cite{Carroll} then
proceeds to obtain the bound $\Lambda < 100$ MeV by requiring the decay
$\phi\to 2g\,3\phi$ to be slower than the present Hubble rate.  But this
bound is model dependent; it requires higher derivative Lagrangian
terms.  It also suffers from the same error of using at-rest decay
rates, so the quoted bound is orders of magnitude too stringent.

\smallskip \centerline{\epsfxsize=0.45\textwidth\epsfbox{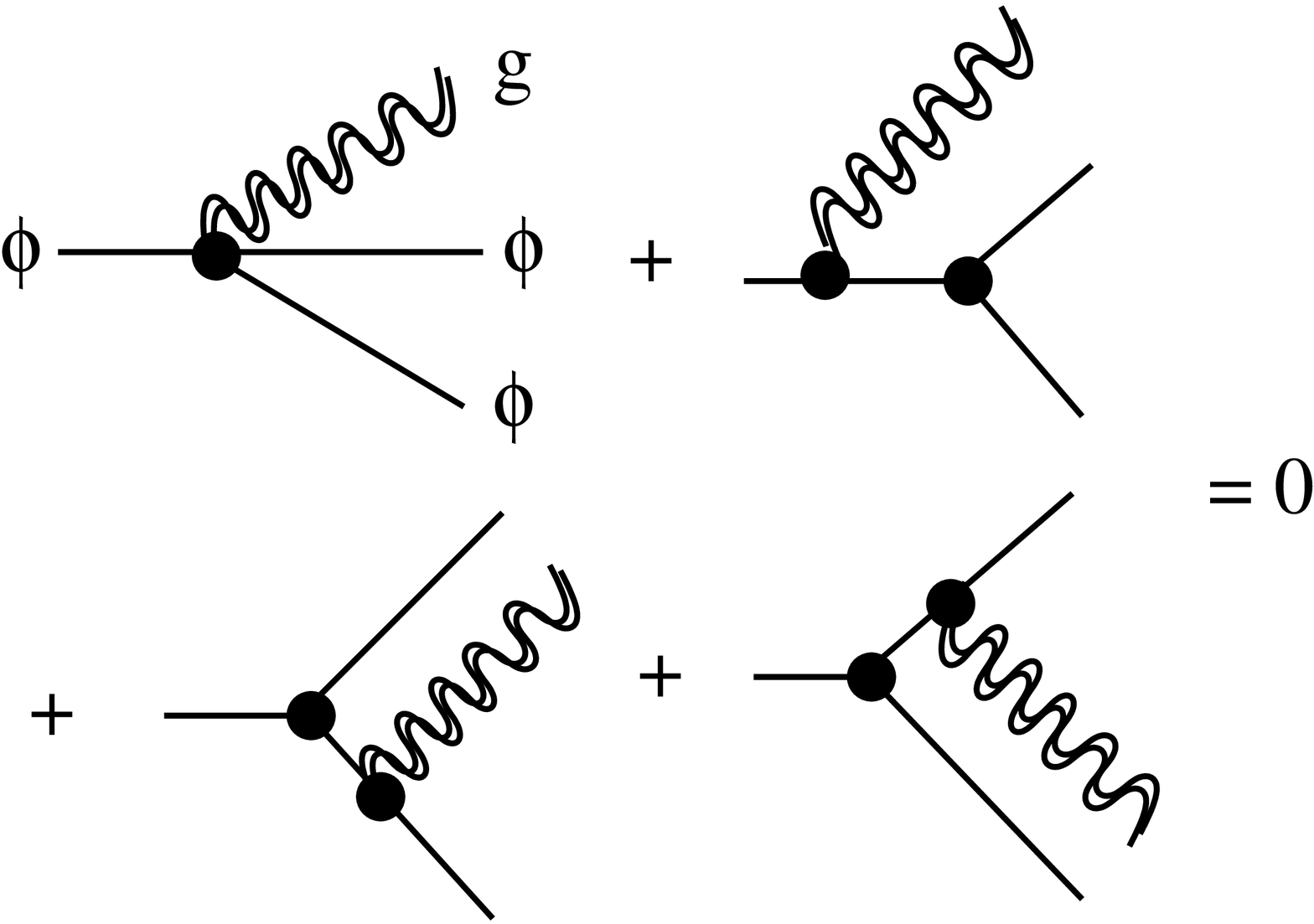}}
\begin{verse}
{\small Figure 2. Vanishing of the amplitude for a  ghost to decay into
graviton and two ghosts.}
\end{verse}


{\it Remarks.} The stringent limit we have obtained from the diffuse gamma ray
background, $\Lambda \lsim 3$ MeV,  implies that any theory of low-energy effective ghosts
must originate from new physics far below the TeV scale.  Therefore we cannot invoke
string theory, for example, as a plausible source for effective ghosts.  Instead, we must
imagine that they come from a low-energy sector that is completely hidden from the
standard model, except for gravitational couplings.  This makes ghosts look even more
unlikely, in our view.

Another troubling feature is that, in order to pose this problem at all,
we were forced to assume that Lorentz symmetry is broken.  By taking the
phase space for production of two photons plus two ghosts to be cut
off at some momentum $\Lambda$, we have singled out a preferred frame,
namely the rest frame of the cosmic microwave background radiation.
Obviously there exist other frames where $k>\Lambda$ even if $k<\Lambda$
in the CMB rest frame.  Such a cutoff might arise if, for example, the ghost
dispersion relation had the form $\omega = -\sqrt{k^2 - k^4/\Lambda^2}$,
which would result from the Lorentz-violating Lagrangian $-\frac12(\partial\phi)^2
+\frac12\Lambda^{-2}(\vec\nabla^2\phi)^2$.
The Lorentz-violating cutoff is necessary because, if we try to
impose a Lorentz {\em invariant} cutoff, for instance on the virtuality
of the off-shell graviton, then there is still a divergent integral over
the boost, with respect to the microwave background frame, of the rest
frame of the (timelike) graviton.

If we demand Lorentz invariance, but want to be maximally conservative,
then we could argue that a process with a formation time longer than the
age of the universe should not be considered.  This places a bound on
the boost between graviton and microwave frames, of $\gamma < t_0
\sqrt{s}$, where $s$ is the Mandelstam variable (the 4-momentum squared of
the off-shell graviton).  Imposing in addition the Lorentz invariant
bound $s < \Lambda^2$ on the graviton propagator, the production rate
becomes finite. Denoting the 4-momentum of the virtual graviton as $k$, 
the production rate is 
\beq
\Gamma \sim \int d^4 k \: \theta(\Lambda^2-k^2) \:
	\theta(t_0-k_0/k^2) \frac{k^4}{M_p^4}
	\sim \frac{\Lambda^{10} t_0^2}{M_p^4} \, ,
\eeq
so that the number density is $\sim \Lambda^{10} t_0^3 M_p^{-4}$.
The typical energy of a produced photon is $k_0 \sim \Lambda^2 t_0$;
even for a cutoff $\Lambda$ of order milli-electron volts, the energy is
$\sim 10^{18}$ GeV.  The dominant mechanism by which gamma rays of such
an energy scatter on the way to the Earth is $\gamma \gamma \to 4e$,
with the second $\gamma$ a microwave background photon; the free path is
about 120 megaparsecs \cite{Protheroe}, leading to about a twenty-fold
reduction in the flux.  Arriving at the Earth, such a gamma ray would
produce an air shower more
energetic than any that have ever been seen.
Using current bounds on the flux of such cosmic rays, less than 1 event
per ${\rm km}^2$ per century \cite{fly_eye}, leads to a constraint 
of $\Lambda \lsim 1$ meV (milli-electron volt).  Gravity would receive order
1 modifications at a length scale $>0.2$ millimeters, in contradiction with
experiment \cite{Adelberger}.  Hence, we conclude that even under very
conservative assumptions, ghosts within a Lorentz invariant framework
are experimentally excluded.

The requirement of Lorentz violation is worrisome, because it is
inconsistent with general covariance.  General covariance is the
framework for general relativity, and it provides the gauge principle
which guarantees the masslessness of the graviton.  There are also
very severe constraints on Lorentz violation within ordinary
particle physics \cite{Lorentz_constraints}; and  Lorentz violation in
another sector tends to be communicated to ordinary particle physics via
graviton loops \cite{cliff}.
It is also troubling that, to our knowledge, no consistent
construction of a low energy theory with ghosts from a ghost-free
fundamental theory exists.

These considerations incline us toward the view that ghosts should not
be feared, not because they are harmless, but because it is very
unlikely that they exist.  The inconveniences of a small cosmological
constant seem much more bearable than those brought on by ghosts.

\bigskip
We thank Nima Arkani-Hamed, Ramy Brustein, Daniel Chung,
 Anne M.\ Green, Justin Khoury, Riccardo Rattazzi and Mark Trodden for useful
remarks.  JC acknowledges the Kavli Institute for Theoretical Physics and CERN theory group
for their hospitality while this work was ongoing.
We are supported in part by the Natural Sciences and
Engineering Research Council of Canada and by le Fonds 
Nature et Technologies of Qu\'ebec.  
S.J.~also
thanks RIKEN-BNL Center and U.S. Department of Energy [DE-AC02-98CH10886] for
providing facilities essential for the completion of this work.

\end{document}